\documentclass[runningheads]{llncs}

\usepackage[T1]{fontenc}
\usepackage{graphicx}
\usepackage{amsmath}
\usepackage{url}
\usepackage[hidelinks]{hyperref}

\title{Poster: Mind the Gap - Characterizing the Temporal Blind Spot Between GSB and DNS Resolution}

\titlerunning{GSB Poster}

\author{Tomer Gal\inst{1} \and
Fujiao Ji
\inst{2} \and
Doowon Kim\inst{2} \and Harel Israel Berger\inst{1}}
\institute{Ariel University \\
\email{harelb@ariel.ac.il,tomergal40@gmail.com}
\and
University of Tennessee, Knoxville \\
\email{fji1@vols.utk.edu,dkim52@utk.edu}}

\begin{document}

\maketitle

\begin{abstract}

Google Safe Browsing (GSB) and DNS resolution operate concurrently during
browser navigation, yet their packet-level synchronization remains understudied. This work characterizes the timing gap (\(\Delta_{time}\)) between GSB-related query close events and parallel DNS resolution responses, identifying a consistent temporal offset with potential security relevance. Using packet-capture analysis across general and CNAME-domain datasets, we observe positive timing gaps in approximately 79\% of measurements. In these instances, DNS responses lag behind GSB-related query closures with median delays of 67--79 ms and maximum delays surpassing 2,400 ms. These empirical results highlight a measurable window within the browsing workflow. We suggest that such temporal inconsistencies, particularly in complex CNAME-domain resolutions, may create a security-relevant timing precondition under DNS-manipulation threat models. These results provide a foundation for further
research into timing-based risks and mitigations in browser safety mechanisms.

\end{abstract}

\section{Introduction}

% Google Safe Browsing (GSB) is widely used by modern browsers and applications to help identify potentially unsafe web resources before users navigate to them. At the same time, web navigation depends on Domain Name System (DNS) resolution, which translates domain names into network destinations. Modern websites often rely on Canonical Name (CNAME) records, where one domain name aliases another and may require additional resolution steps before the final address is obtained.
Modern web navigation relies on the parallel execution of browser-level safety
checks and DNS address resolution. While Google Safe Browsing
(GSB)~\footnote{https://safebrowsing.google.com/} helps identify potentially malicious resources, DNS maps domains to network destinations. This workflow becomes more complex when domains resolve through Canonical Name (CNAME) chains, which may require multiple DNS steps before the final address is obtained.

Although both GSB~\cite{Drury2022,Han2016Phisheye,Lee2025WWW,Adam2020Sunrise,Rana2024AsiaCCS}
and DNS~\cite{Afek2025POPS,Li2025RebirthDay,son2010hitchhiker}
have been studied as independent attack surfaces, their packet-level synchronization within the browser workflow remains underexplored. This work examines whether GSB-related query close events and final DNS responses exhibit a measurable temporal gap with potential security relevance, particularly for CNAME-based resolutions.

We evaluate this relationship using general-domain and CNAME-domain datasets. Positive timing gaps appear in approximately 78\%--80\% of measured samples, with median delays around 70~ms and maximum observed delays exceeding two
seconds.

Our study makes the following contributions:
\begin{enumerate}
  \item We define a packet-level timing metric for comparing GSB-related query close events and final DNS response events.
  \item We measure this relationship across general-domain and CNAME-domain datasets.
  \item We quantify the prevalence and magnitude of positive timing gaps across the measured datasets.
\end{enumerate}

% This work focuses strictly on empirical timing measurements and does not infer exploitability, security impact, response validity, or correctness of the underlying responses.
\section{Background}

\subsection{Google Safe Browsing and DNS Resolution}

Google Safe Browsing (GSB) helps identify potentially unsafe web resources
during browser navigation. To balance performance and privacy, browsers
typically check a local hash-prefix database and query GSB servers only when a
prefix match occurs. This work does not evaluate GSB's detection accuracy;
rather, it studies the packet-level timing of GSB-related connection events
relative to DNS resolution.

DNS translates domain names into network-level addresses. While a standard
lookup may directly return an A or AAAA record, modern web infrastructure often
uses Canonical Name (CNAME) records, where one domain aliases another before a
final address is obtained. Such multi-step resolution can introduce additional
latency, motivating our comparison between general-domain and CNAME-domain
measurements.

\subsection{Threat Model and Security Relevance}

The security concern arises because Safe Browsing-related checks and DNS
resolution may execute concurrently. If the GSB-related check completes before
the final DNS or CNAME result is available, the browser may reach a safety
decision before the final network endpoint is fully established.

We consider an adversary who can influence DNS resolution, for example through
cache poisoning, resolver manipulation, or an on-path position. Under this
threat model, a positive timing gap may allow the final resolved endpoint to be
affected after the Safe Browsing-related check has completed. We do not claim a
complete browser bypass; instead, we identify a timing precondition that may
enable a mismatch between the Safe Browsing decision and the final DNS-resolved
endpoint.
\section{Methodology}

\subsection{Metric and Extraction}
\label{metric}

To characterize the timing relationship between Google Safe Browsing
(GSB)-related connection events and DNS resolution, we define the timing gap as:

\begin{equation}
\Delta_{\mathrm{time}} = T_{\mathrm{resolution}} - T_{\mathrm{query\_close}}
\end{equation}

where \(T_{\mathrm{query\_close}}\) represents the transport-layer closure of
the GSB-related connection and \(T_{\mathrm{resolution}}\) is the arrival of the
final DNS response. A positive \(\Delta_{\mathrm{time}}\) indicates a measurable
timing window in which the final DNS response is observed after the
GSB-related query close event.

We identify GSB traffic via TLS Server Name Indication (SNI) metadata. Since
the application data is encrypted, we define \(T_{\mathrm{query\_close}}\) as
the timestamp of the associated TCP FIN or RST packet. To ensure a clear
termination point for measurement, our automated navigation was configured to
treat the transport-layer closure (FIN/RST) of the GSB-related stream as the
query close event. For CNAME-based resolutions, \(T_{\mathrm{resolution}}\) is
recorded from the packet containing the final A/AAAA record in the resolution
chain.

\subsection{Data Selection}

We evaluated this metric across two distinct datasets derived from the Tranco
rank~\cite{pochat2018tranco}:

\begin{itemize}
    \item \textbf{General Domains:} The top 100 most popular domains, used to
    establish a baseline for standard navigation.
    \item \textbf{CNAME Domains:} 103 domains identified by scanning the Tranco
    list for the highest-ranked entries that explicitly require CNAME resolution.
\end{itemize}

% The packet timestamps are recorded in seconds, while the computed timing gap is reported in milliseconds. This methodology is used only to measure event ordering and timing differences; it does not infer exploitability, security impact, response validity, or correctness of the underlying responses.
\section{Experimental Design}

Measurements were conducted on a macOS workstation using Google Chrome Stable,
selected because of its native integration with Google Safe Browsing. Before
each navigation, we cleared the system-level DNS cache and browser state to
reduce the effect of cached records and persistent sessions.

Our Node.js-based framework coordinated browser automation and packet capture.
For each target domain, the pipeline reset the environment, started
\texttt{Tshark} to capture DNS and HTTPS traffic, launched an ephemeral Chrome
instance to navigate to the target URL, and post-processed the resulting trace
to extract \(T_{\mathrm{query\_close}}\) and \(T_{\mathrm{resolution}}\)
according to Section~\ref{metric}.
\section{Results}

Table~\ref{tab:per_domain_timing} presents representative per-domain timing measurements. For each domain, we report the GSB-related query close timestamp, the corresponding final DNS response timestamp, and the measured gap between the two events.The packet-capture timestamps are reported in seconds, while the computed timing gap is reported in milliseconds.

\subsection{Observed Timing Gaps}

The results show that the final DNS response is observed after the GSB-related query close event for all domains listed in Table~\ref{tab:per_domain_timing}. However, the size of the measured timing gap is not uniform across domains. The smallest measured gap is observed for \textit{facebook.com}, with a gap of
45~ms, followed by \textit{live.com} with 61~ms. Similar timing gaps are observed for \textit{youtube.com} and \textit{microsoft.com}, with gaps of 118~ms and 119~ms,respectively.

Larger timing differences are observed for several domains. \textit{apple.com}
and \textit{instagram.com} show gaps of 189~ms and 267~ms, respectively, while
\textit{office.com} and \textit{twitter.com} show larger gaps of 403~ms and 566~ms. The largest timing difference in this representative set is observed for \textit{amazonaws.com}, where the final DNS response is observed 1034~ms after the GSB-related query close event. These measurements indicate a measurable temporal window in which DNS resolution may complete after the
GSB-related query close event.

\subsection{Dataset-Wide Analysis}

Table~\ref{tab:dataset_summary} summarizes the positive timing gaps across the
two measured datasets. The general-domain dataset contains 100 samples, of which 78 show a positive timing gap. The CNAME-domain dataset contains 103 samples, of which 82 show a positive timing gap. The dataset-level results show that positive timing gaps appear in a large fraction of the measured samples: 78.0\% in the general-domain dataset and 79.6\% in the CNAME-domain dataset. Although many measured gaps are below 100~ms, the maximum observed gaps exceed two seconds in both datasets. Overall,
these measurements indicate that the timing gap appears consistently across the representative sample and across the larger datasets, while its magnitude varies substantially across domains. The observed long-tail delays motivate further security analysis, especially for DNS resolution patterns involving complex
CNAME chains.

\begin{table}[h]
\caption{Representative per-domain timing differences}
\label{tab:per_domain_timing}
\centering
\begin{tabular}{l r r r}
\hline
\textbf{Domain} &
\textbf{Query Close Time (s)} &
\textbf{Response Time (s)} &
\textbf{Gap (ms)} \\
\hline
microsoft.com & 6.847 & 6.966 & 119 \\
facebook.com  & 1.744 & 1.789 & 45 \\
amazonaws.com & 1.426 & 2.460 & 1034 \\
youtube.com   & 1.786 & 1.904 & 118 \\
apple.com     & 1.647 & 1.836 & 189 \\
instagram.com & 1.621 & 1.888 & 267 \\
office.com    & 1.575 & 1.978 & 403 \\
twitter.com   & 1.604 & 2.170 & 566 \\
live.com      & 1.698 & 1.759 & 61 \\
\hline
\end{tabular}
\end{table}

\begin{table}[h]
\caption{Dataset-level summary of positive timing gaps}
\label{tab:dataset_summary}
\centering
\begin{tabular}{l r r r r r}
\hline
\textbf{Dataset} &
\textbf{N} &
\textbf{Positive Gaps} &
\textbf{Mean (ms)} &
\textbf{Median (ms)} &
\textbf{Max (ms)} \\
\hline
General & 100 & 78 (78.0\%) & 248.12 & 79.40 & 2349.54 \\
CNAME   & 103 & 82 (79.6\%) & 178.79 & 67.88 & 2477.32 \\
\hline
\end{tabular}
\end{table}

\section{Security Implications}

The observed timing gaps, reaching up to 2.4 seconds, suggest a
security-relevant window that DNS-manipulation attacks could potentially
exploit. In this window, a Safe Browsing-related decision may complete before
DNS resolution has fully determined the final network endpoint. Prior DNS
attacks have shown that resolution outcomes may be manipulated within short
time windows~\cite{li2024tudoor}. Our findings therefore indicate a possible
attack precondition: if an adversary can influence DNS resolution, for example
through cache poisoning, resolver manipulation, or an on-path position, the
browser may reach a final endpoint that is determined only after the initial
Safe Browsing-related check has completed. We do not claim a complete browser
bypass, but identify a measurable timing mismatch that motivates further
end-to-end validation.

\section{Conclusion}

This study provides a packet-level characterization of the temporal
desynchronization between Google Safe Browsing (GSB)-related checks and DNS
resolution. Our empirical analysis of general and CNAME-intensive domains
reveals that a positive timing gap is a recurring feature of the measured
browsing workflow, occurring in approximately 80\% of all cases.
While the median gap remains below 100~ms, the presence of outliers exceeding
two seconds shows that DNS resolution may continue after the Safe
Browsing-related check has completed. These results do not establish a complete
browser bypass. Instead, they identify a timing precondition that may enable a
mismatch between the browser's safety decision and the final DNS-resolved
endpoint under DNS-manipulation threat models.

Future work will expand this study across a broader range of network vantage
points, operating systems, and browsers, including Chromium-based and
non-Chromium browsers. We also plan to evaluate controlled DNS-manipulation
scenarios and longer CNAME chains to determine whether the measured timing gaps
can be converted into a practical end-to-end attack. Finally, we will explore
browser-side mitigation strategies, such as binding Safe Browsing decisions to
the final resolved DNS endpoint or triggering revalidation when DNS resolution
completes after the initial safety check.

\bibliographystyle{plain}
\bibliography{biblio}

\end{document}